\documentclass[aps,pra,twocolumn,groupedaddress,superscriptaddress]{revtex4}
\usepackage{amsmath}\usepackage{xcolor}\usepackage{epstopdf}
\usepackage{bbold}
\usepackage{graphicx}
\usepackage{color,ulem}

\usepackage{xcolor}

\newcommand{\be}{\begin{equation}}
\newcommand{\ee}{\end{equation}}
\newcommand{\beq}{\begin{eqnarray}}
\newcommand{\eeq}{\end{eqnarray}}
\newcommand{\bea}{\begin{array}}
\newcommand{\eea}{\end{array}}

\usepackage{hyperref}
\usepackage{xcolor}
\usepackage{color}
\usepackage{amsmath}
\usepackage{ulem}
\usepackage{gensymb}
\usepackage{amsmath}
\usepackage{ulem}
\usepackage{gensymb}

\draft % marks overfull lines with a black rule on the right
\begin{document}
\title{Tailoring radiation pressure on infinite slab using pair of non-collinear plane waves } %Title of paper
\author{R. Ali}
\email[]{rali.physicist@gmail.com}

\affiliation{%
Applied Physics Department, Gleb Wataghin Physics Institute, University of Campinas, Campinas 13083-859, SP,
Brazil
}

\author{R. S. Dutra}
%\email[]{rali.physicist@gmail.com}

\affiliation{%
{LISComp-IFRJ },  
{Instituto Federal de Educa\c c\~ao, Ci\^encia e Tecnologia, Rua Sebasti\~ao de Lacerda, Paracambi, RJ, 26600-000, Brasil }
}

\author{S. Iqbal}
%\email[]{rali.physicist@gmail.com}

\affiliation{%
{Department of Physics},  
School of Natural Sciences
National University of Sciences and Technology
Sector H-12, Islamabad, Pakistan}

\date{\today}

\begin{abstract}
The electromagnetic field exerts radiation pressure on the matter and tends to move it either in the backward or forward direction due to net optical pulling or pushing force, respectively. In this work, we reveal an interesting phenomenon of a local positive {and negative } radiation pressure on { a} dielectric  (chiral)  slab by using two linearly {(circularly)} polarized plane waves. { In this regard}, we develop for the first time, a theory to describes the local radiation pressure { appearing} due to the interference between the two obliquely impinging (non-collinear) light sources. Under this {situation}, the radiation pressure strongly depends on the angle of incidence, the polarization of the electromagnetic { field} and {the} chirality parameters {of the slab (in the case of chiral medium)}. {Our numerical} analysis shows that the radiation pressure{,} exerted on a dielectric { or a} chiral slab due to {the} two incident plane waves{,} are constant over the slab for normal incidence{,} and {it varies locally for an oblique incidence}, which indeed follows the {conservation} laws { at} all {incident} angles.  It is expected that the results may find fruitful applications in optical manipulation of soft matters, {for instance,} cell membranes, chiral surfaces and other soft materials.
\end{abstract}
\maketitle
\section{Introduction}
The electromagnetic field carries both linear and angular momentum which can be exchanged during the field-matter interaction, and yield a radiation pressure and radiation torque  on the matter \cite{Ashkin1970}. 
In the past few { decades}, the field of light-matter interaction has grown impressively after the invention of optical tweezers \cite{ashkin1986} that have played an indispensable role with countless applications \cite{Gieseler2020,Polimeno2018}. In optical tweezers setups, the trapping and manipulating of spherical objects are obtained near to the focus {point} of a tightly focused continuous-wave laser beam \cite{ashkin2006,Friese1998,grier2003,Lin2014}, while the long-range optical manipulation of  chiral and dielectric objects can be achieved by using a structured Bessel beam or two non-collinear plane waves  \cite{Ali2020Tailor,Ali2021,chan2011,li2019,Shvedov2014,dogariu2011}. 

The theory of optical manipulation is not limited to spherical  particles but  tremendous efforts  have been made to calculate the radiation pressure on various kinds of objects, such as, composite spheres \cite{Ali2020Tailor,Ali2018}, spheroidal particles  \cite{Stephen2007}, nanorices \cite{Wiley2007}, cylindrical particles \cite{Gauthier1999}, chiral spheres \cite{Ali2020,Ali2020josa,Canaguier2013,Kun2014}, chiral nanocrystals \cite{Ali2020} and dielectric slabs \cite{Mansuripur2005,Loudon2010,Alaee2018},   wherein the magnitude as well as direction of the optical { forces} and the flow of optical  momenta  have been discussed.
Nowadays, metamaterials and metasurfaces are emerging fields that have potential applications in the field of nanotechnology, sensing and medicine \cite{Corradini2007}. The negative radiation pressure on such surfaces, for instance, cell  membranes, and chiral metasurfaces \cite{Basiri} could be used to probe the local properties  by stretching  and relaxing the surfaces due to the incident light. To the best of our knowledge, the local radiation pressure on such metasurfaces has not been discussed so far. 

It is worth mentioning that a single plane wave is obliged to exert positive radiation pressure on any passive system in the paraxial approximation due to momentum conservation \cite{Lepeshov2020,Ali2020,ashkin2006}. For example,  the  optical force $F$ acting on a  slab located in vacuum can be expressed as  $F =N_i\hbar k {(1 + r - t)}$, where $N_i = c \varepsilon_0  E_0^2/2\hbar \omega$ is the photon density, $E_0^2$ is the electric field intensity, c is the speed of light, $\omega$ is the angular frequency of the light and $\varepsilon_0$ is the vacuum permittivity. Furthermore,  {$t $ and $r$} are the Fresnel transmission and reflection coefficients that use to define the momentum transmitted and reflected through the slab,  respectively. In this regard,  the optical force exerted on a  \textit{perfect reflector } ($ r = 1, \, t = 0$),  \textit{black body}  ($ { r} = 0,\, t = 0$), and    \textit{perfect transmitter} ($ r = 0, \, t  = 1$) can be  estimated as  $F = 2 N_i\hbar k$, $F = N_i\hbar k $ and  $F = 0 $, respectively.  Thus, the net  force  can not be negative due to fact that  the  transmitted or reflected  photons cannot have larger momentum than the incident ones and hence, $F\geq  0$ \cite{Lepeshov2020,Mizrahi2010,Wang2015,Li2016}.
 
  However, the situation  changes dramatically when we employ optical gain \cite{Ali2021gain} or  shine (normally)  a structured incident beam such as Bessel beam. In { such} cases, one can tune the direction of scattered momentum flow in forward direction and due to conservation of  momentum, the recoiling force could be negative. In all the cases, the optical force or indeed the radiation pressure is constant over the slab, thereby, net force appears at the center of  mass of the slab.
 
In this paper, we are intended  to calculate the locally negative and positive radiation pressure density exerted {by a pair of } plane waves {on the slab by evaluating the Maxwell stress tensor}. To this end, we consider { an} oblique incidence of two linearly polarized plane waves separated by an angular distance $2\theta$ (see Sec. { \ref{method}} ) impinging on the dielectric slab. {Furthermore,  we  extend this framework to calculate the negative radiation pressure { exerted by two circularly polarized non-collinear plane waves} on } chiral { slabs, for instance,  slabs wherein} the chirality may appear due to randomness, chiral metasurfaces \cite{Basiri} or naturally {occurring biological  surfaces}. In all the cases, we examine the radiation pressure acting {on} the slab due to interference of incident plane waves and analyze the role of incident angle $\theta$ and chirality { parameter}.  In this perspective, we presented for the first, a new analytical framework to calculate the momentum density over the slab due to the superposition of plane waves {of} light, which meets all {the} necessary conservation laws for all {the incident} angles. 

The rest of the paper is organized as follows. Section {\ref{method} } is devoted to the derivation of our theoretical formalism, where we derive the optical radiation pressure expressions for dielectric and chiral slab and discussed its limiting cases to well-known results.  In Sec. \ref{results}, we present numerical results and analyze the findings. Finally, we summarize our findings and conclusions in Sec \ref{discussion}.
\section{Methodology} \label{method}

\subsection{Radiation pressure on dielectric slab due to two linearly polarized plane waves}

\begin{figure*}[hbt!]
	\centering
	\includegraphics[scale=0.5]{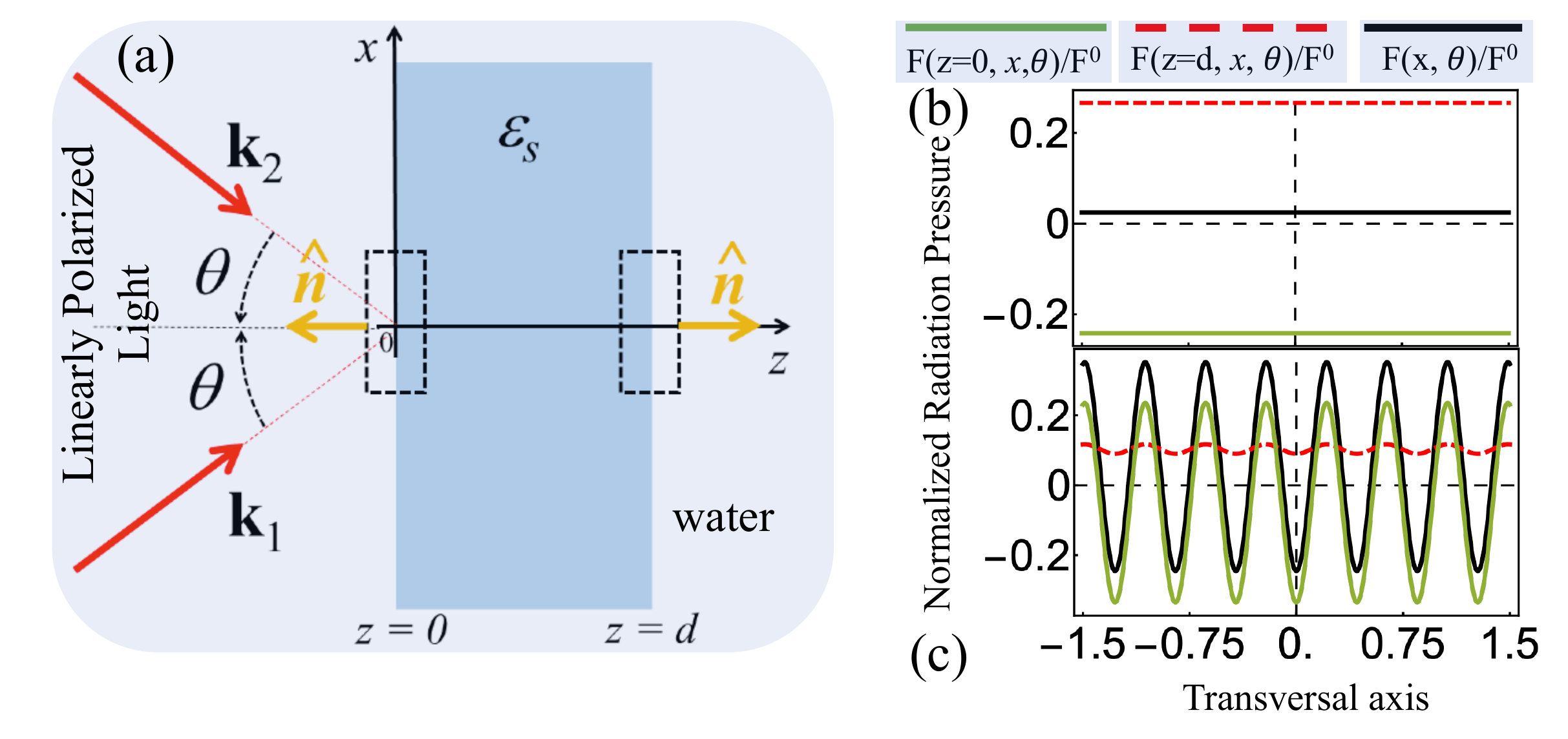}
	\caption{{Schematic diagram: an infinite  dielectric slab of thickness $d$, relative permittivity $\epsilon_s$ is  illuminated by two
non-collinear transverse electric plane waves making angle $
\theta$  with the z-axis, where dashed lines show the surfaces of the imaginary Gaussian surfaces with normal unitary vector $\hat{n}.$} { Radiation pressure acting on the slab interface $z=0$ (green) and  $z=d$ (dotted) and total local radiation pressure on  $F({x,\theta})= F({z=0,x,\theta})+F({z=d,x,\theta})$ (black) for (b) normal incidence and (c) oblique incidence at $\theta=70^o$. }}
	\label{F1} 
\end{figure*}

We consider two linearly polarized transverse electric (TE) plane waves impinging obliquely at an angle $\theta_i$ (where $i= 1, 2$) on an infinite  dielectric slab of thickness $d$ and refractive index $n_s=\sqrt{\varepsilon_s \mu_s}$, where $\varepsilon_s$ and $\mu_s$ are the relative  permittivity and relative permeability of the slab, as depicted in figure \ref{F1}. 
The total  impinging  electric field on the slab is defined by taking the superposition of both plane waves propagating through the host medium of  relative permittivity $\varepsilon$  as 

\begin{equation}
\mathbf{E}_{in,total}=E_{0}\sum_{j=1}^2e^{i \mathbf{k}_{j}\cdot\mathbf{r}}\hat{y}, \label{Einp}
\end{equation}
where $\mathbf{k}_j$ ($j= 1, 2$)  is the  wave vector of $j^{th}$ plane wave and can be expressed  as  $\mathbf{k}_{j}=(-1)^{j+1}k\sin\theta_{j}\mathbf{\hat{x}}+k\cos\theta_{j}\mathbf{\hat{z}}$. {At the boundary of the slab, a part of $\mathbf{E}_{in,total}$ will be reflected back into surrounding medium and} 
propagate with  wave vector $\mathbf{k}'_{j}=(-1)^{j+1}k\sin\theta\mathbf{\hat{x}}-k\cos\theta\mathbf{\hat{z}}$, for the sake of easiness  we take  $\theta_{1}=\theta_{2}=\theta$. By conserving the polarization, the total reflected field $\mathbf{E}_{r,total}$ can be written  as 

\begin{equation}
\mathbf{E}_{r,total}=E_{0}\,r_{TE}\sum_{j=1}^2 e^{i \mathbf{k}'_{j}\cdot\mathbf{r}}\hat{y}, \label{Erp}
\end{equation}
where $r_{TE}$ is reflection  coefficient for infinite slab \cite{Born}.  
The rest part of  $\mathbf{E}_{in,total}$ is transmitted through the slab at $z=d$ with the same incident angle $\theta$ and allowing us to write the total transmitted electric field  as 

\begin{equation}
\mathbf{E}_{t,total}=E_{0}\,t_{TE}\sum_{j=1}^2 e^{i \mathbf{k}_{j}\cdot\mathbf{r}}\hat{y}, \label{Etp}
\end{equation}
where, $t_{TE}$  is the transmission coefficient of  an infinity slab \cite{Born}. 

Likewise, the similar expressions for the total incident, reflected and transmitted magnetic fields can also be obtained by using the relation $\mathbf{H}_{j}=\frac{\mathbf{k}_{j}}{k_{j}\mu v}\times \mathbf{E}_{j}$  for each plane wave, where $\mu$ and $v$ are the {host medium} magnetic permeability and the speed of light,  respectively. 

The radiation pressure on an infinite  slab is calculated by using the Maxwell stress tensor $\overleftrightarrow{T}$ \cite{Brandon} through the closed  Gaussian  surfaces as represented by the dashed lines in Fig. \ref{F1} \cite{Brandon} on the both sides of the slab, such as $z=0^{-}$ and $z=d^{+}$, with respective normal unitary vectors $\mathbf{\hat{n}}=\mathbf{-\hat{z}}$ and $\mathbf{\hat{n}}=\mathbf{\hat{z}}$, respectively. 
Finally, the radiation pressure $\mathbf{F}$ can be defined as
\begin{equation}
\mathbf{F} = \frac{1}{2}\Re\lbrace \mathbf{\hat{z}}\cdot \overleftrightarrow{T}(z=0^{-})- \mathbf{\hat{z}}\cdot\overleftrightarrow{T}(z=d^{+})\rbrace, \label{pressure}
\end{equation}
 where $\overleftrightarrow{T}(z = 0)$ and  $\overleftrightarrow{T}(z = d^+)$   are the Maxwell stress tensor  evaluated at point z =0 and  z=d. Without going to the debate on the Maxwell stress tensors,  here we are using Minkowski Maxwell stress tensor that take into account the medium contribution \cite{Pfeifer2007}.
The explicit expression for $\overleftrightarrow{T}$ can be found by substituting the total electric and magnetic fields in the standard expression (see Eq. 4 of the Ref. \cite{Brandon})  of  the Maxwell stress tensor and is expressed as    

\begin{eqnarray}
\overleftrightarrow{T}= \frac{I}{2}(\varepsilon\mathbf{E}_{total}\cdot\mathbf{E}^{*}_{total}+\mu \mathbf{H}_{total}\cdot\mathbf{H}^{*}_{total})\nonumber\\-\varepsilon\mathbf{E}_{total}\mathbf{E}^{*}_{total}-\mu\mathbf{H}^{*}_{total}\mathbf{H}_{total} \label{tensor}
\end{eqnarray}
{where $\mathbf{E}_{total}(z=0^{-})=\mathbf{E}_{in,total}(z=0^{-})+\mathbf{E}_{r,total}(z=0^{-})$,\,  $\mathbf{E}_{total}(z=d^{+})=\mathbf{E}_{t,total}(z=d^{+})$ and {  $\varepsilon$ is  relative permittivity of the surrounding medium. }}
By setting  the Eqs. (\ref{Einp})-(\ref{Etp}),
we obtain the total radiation pressure on the slab by two non-collinear plane waves 

\begin{eqnarray}
&&\mathbf{F}=\mathbf{F}(x,\theta)= \varepsilon E_{0}^2\biggl[\biggl(\vert 1+r_{TE}\vert^2+\vert 1 - r_{TE}\vert^2 \cos^2\theta\nonumber\\&&-\vert t_{TE} \vert^2( 1+cos^2\theta)\biggr)\,\cos^2(kx\sin\theta) {+}
\biggl(\vert t_{TE} \vert^2 - \vert 1+r_{TE} \vert^2\biggl)\nonumber\\&&\sin^2\theta\,\sin^2(kx\sin\theta)\biggr]\mathbf{\hat{z}}. \label{force_linear}
\end{eqnarray}
Since the problem involves the electromagnetic field solutions inside and outside the slab and the transmission $t_{TE}$ and reflection $r_{TE}$ coefficients are obtained by applying the appropriate  boundary conditions at the interfaces between  the surrounding medium and the slab. The final expression are given as

\begin{equation}
r_{TE} =\frac{r_{m,s}+r_{s,m} exp(i 2\beta)}{1+r_{m,s}r_{s,m} exp(i 2\beta)} \label{rslab}
\end{equation} 
and

\begin{equation}
t_{TE} =\frac{t_{m,s}t_{s,m} exp(i \beta)}{1+r_{m,s}r_{s,m} exp(i 2\beta)}, \label{tslab}
\end{equation}

where, $\beta = 2 \pi n_s d \cos\theta/\lambda$  is optical path inside the slab. The coefficients  ($t_{m,s}$, $r_{m,s}$) and ($t_{s,m}$, $r_{s,m}$) are evaluated  at the interfaces and are expressed as 
\begin{equation}
t_{m,s} = \frac{2 n_m \cos\theta_m}{n_m \cos\theta_m+ n_s \cos\theta_s},
\end{equation}

\begin{equation}
t_{s,m} = \frac{2 n_s \cos\theta_s}{n_s \cos\theta_s+ n_m \cos\theta_m}
\end{equation}

and 
\begin{equation}
r_{m,s} = -r_{s,m}=\frac{n_m \cos\theta_m -n_s  \cos\theta_s}{n_m \cos\theta_m+n_s \cos\theta_s},
\end{equation} 
 The refraction angle $\theta_s$ inside slab and the incident angle $\theta_{m}=\theta$  are connected by Snell's law $\theta_s=\arcsin({n_m\sin\theta/n_s})$.

In order to gain a more physical insight,
we compare the limiting case of our exact analytical expressions presented in Eq. \ref{force_linear}, when $\theta=0$  we recover the well known results for the radiation pressure acting on an infinite slab  with gain/loss due to a normal incidence  \cite{Alaee2018,Lepeshov2020,Mizrahi2010,Brandon}.

\begin{equation}
\mathbf{F}{{(0\degree)}}= 2\varepsilon E_{0}^2\biggl[1+\vert r_{TE} \vert^2- \vert t_{TE} \vert^2 \biggl]\mathbf{\hat{z}}. \label{F_norm}
\end{equation}

{ It is clear from Eq. \ref{F_norm}, the radiation pressure due to normal incidence does not depend on transversal position. Thus, the pressure obliged to be  constant on each interface of the slab Fig.\ref{F1}(b).
However, when both plane waves are impinging obliquely then the reflected and transmitted light are expected to  produce an interference pattern along x-direction. Thus, each surface of the  slab undergoes to the local maximum and minimum  radiation pressure as shown in Fig.\ref{F1}(c). In Fig.\ref{F1}(b)-(c)  we calculate the radiation pressure on each interface for fixed thickness $d=\lambda$, $n_m=1.332$ and $n_s=1.5$.  The details of analytical calculation of the radiation pressure acting on interfaces $z=0:$ $\mathbf{F}(z=0, x,\theta)$ and $z=d:$  $\mathbf{F}(z=d, x,\theta)$  are presented in appendix A.  }

\subsection{Radiation pressure on chiral slab due to two circularly  polarized plane waves}

Propagation of the electromagnetic wave  through a chiral media can be  explained by the following constitutive relations 
that connect the complex displacement field $\textbf{D}$ and magnetic field  $ \textbf{B}$ \cite{Ali2020,Ali2020josa,wang2014,Bohren,Ali2020Theory} as 
\begin{gather} 
   \begin{bmatrix} 
 {\bf D}\\  {\bf B} \end{bmatrix}
 = 
  \begin{bmatrix} \varepsilon_0  \varepsilon & i \sqrt{\varepsilon_0 \mu_0}  \, \kappa\\
   -i\sqrt{\varepsilon_0 \mu_0}  \,\kappa&\mu_0 \mu 
  \end{bmatrix} \begin{bmatrix}
   {\bf E} \\
   {\bf H}  
   \end{bmatrix}, \label{constitutive_relation}
   \end{gather} 
  through  a direct  coupling between the electric field $ {\bf  E}$ and the auxiliary field  $ {\bf H} $ by  chirality parameter $ \kappa$.
Here, $\varepsilon$ and $\mu$ are the relative permittivity and relative permeability of the medium,  respectively. The constant  $\kappa$  characterizes the nature of media and usually satisfies the condition $\kappa\ll \sqrt{\varepsilon \mu}$. 
By using the aforementioned constitutive relations in the  Maxwell equations \cite{Bohren}, one can derive the insightful information such that the each circular polarization field propagates with different phase velocity and wave vectors that explicitly depends on chiral parameter $\kappa$. The wave vectors inside the chiral media can be written as  $k_{s,\sigma}=k_{0}(\sqrt{\epsilon_{s}}+\sigma\kappa)$, where $k_{0}= 2\pi c/\lambda_0$ the vacuum wave vector, $\sigma = 1 (-1)$  for left (right) circularly polarized plane wave.  

\begin{figure}[hbt!]
	\centering
	\includegraphics[scale=0.4]{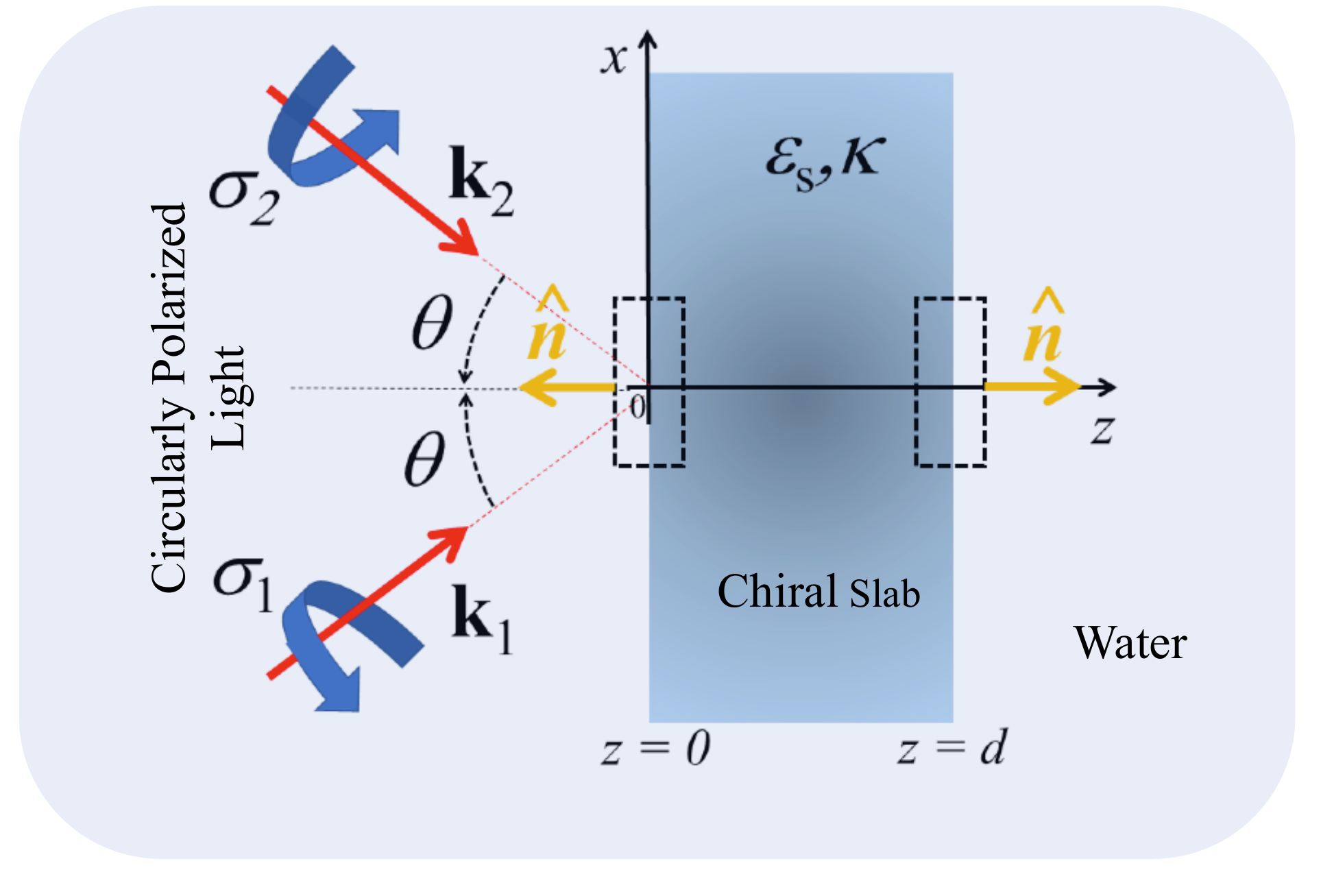}
	\caption{{Schematic illustration  of  the system, an infinite slab of thickness $d$, relative permittivity $\epsilon_s$, chirality parameter $\kappa$ illuminated by two non-collinear circularly polarized plane waves impinging at an angle $\theta$ with
respect to the z axis, $\hat{n}$ is the normal  unitary vector. } 
}
	\label{F2} 
\end{figure}

In this section, we consider two oblique circularly polarized incident plane waves impinging on an infinite  chiral slab, immersed in water as depicted in Fig. \ref{F2}. Following the same procedure, as adopted in the previous section, one can write the total incident field as 

\begin{equation}
\mathbf{E}_{in,total}=E_{0}\sum_{j=1}^2 e^{i \mathbf{k}_{j}\cdot\mathbf{r}}\frac{(\mathbf{\hat{p}}_{j}+i\sigma_{j}\mathbf{\hat{s}}_{j})}{\sqrt{2}}, \label{Ein}
\end{equation}
with unitary vectors $\mathbf{\hat{p}}_{j}= cos\,\theta_{j}\,\mathbf{\hat{x}}+(-1)^j\,sin\,\theta_{j}\,\mathbf{\hat{z}}$ and $\mathbf{\hat{s}}_{j}=\mathbf{\hat{y}}$. The plane waves impinge the slab surface with wave vectors $\mathbf{k}_{j}=(-1)^{j+1}k\sin\theta_{j}\mathbf{\hat{x}}+k\cos\theta_{j}\mathbf{\hat{z}}$, and we assume,  $\theta_{1}=\theta_{2}=\theta$, for simplicity.

The reflected plane waves that propagate with the wave vectors $\mathbf{k}'_{j}=(-1)^{j+1}k\sin\theta_{j}\mathbf{\hat{x}}-k\cos\theta_{j}\mathbf{\hat{z}}$ are expanded in terms of the linearly polarized  basis as

\begin{equation}
\mathbf{E}_{r,total}=\sum_{j=1}^2 e^{i \mathbf{k}'_{j}\cdot\mathbf{r}}\frac{(E_{r,\parallel}\mathbf{\hat{p}}'_{j}+E_{r,\perp}\mathbf{\hat{s}}'_{j})}{\sqrt{2}}, \label{Er} 
\end{equation}
with the reflected unit vectors $\mathbf{\hat{p}}'_{j}= -\cos\theta_{j}\,\mathbf{\hat{x}}+(-1)^j\,\sin\theta_{j}\,\mathbf{\hat{z}}$ and $\mathbf{\hat{s}}'_{j}=\mathbf{\hat{s}}_{j}=\mathbf{\hat{y}}$.

After crossing the slab, the plane waves are transmitted to dielectric host medium, with the same incident angles $\theta_{1}=\theta_{2}=\theta$. We also expand the transmitted total electric field on the linearly polarized basis as

\begin{equation}
\mathbf{E}_{t,total}=\sum_{j=1}^2 e^{i \mathbf{k}_{j}\cdot\mathbf{r}}\frac{(E_{t,\parallel}\mathbf{\hat{p}}_{j}+E_{t,\perp}\mathbf{\hat{s}}_{j})}{\sqrt{2}}. \label{Et} 
\end{equation}
The analogous expressions for the total incident, reflected and transmitted magnetic fields can be obtained in a trivial  way by using the relation $\mathbf{H}_{j}=\frac{\mathbf{k}_{j}}{k_{j}\mu v}\times \mathbf{E}_{j}$.

\begin{figure}[hbt!]
	\centering
	\includegraphics[scale=0.32]{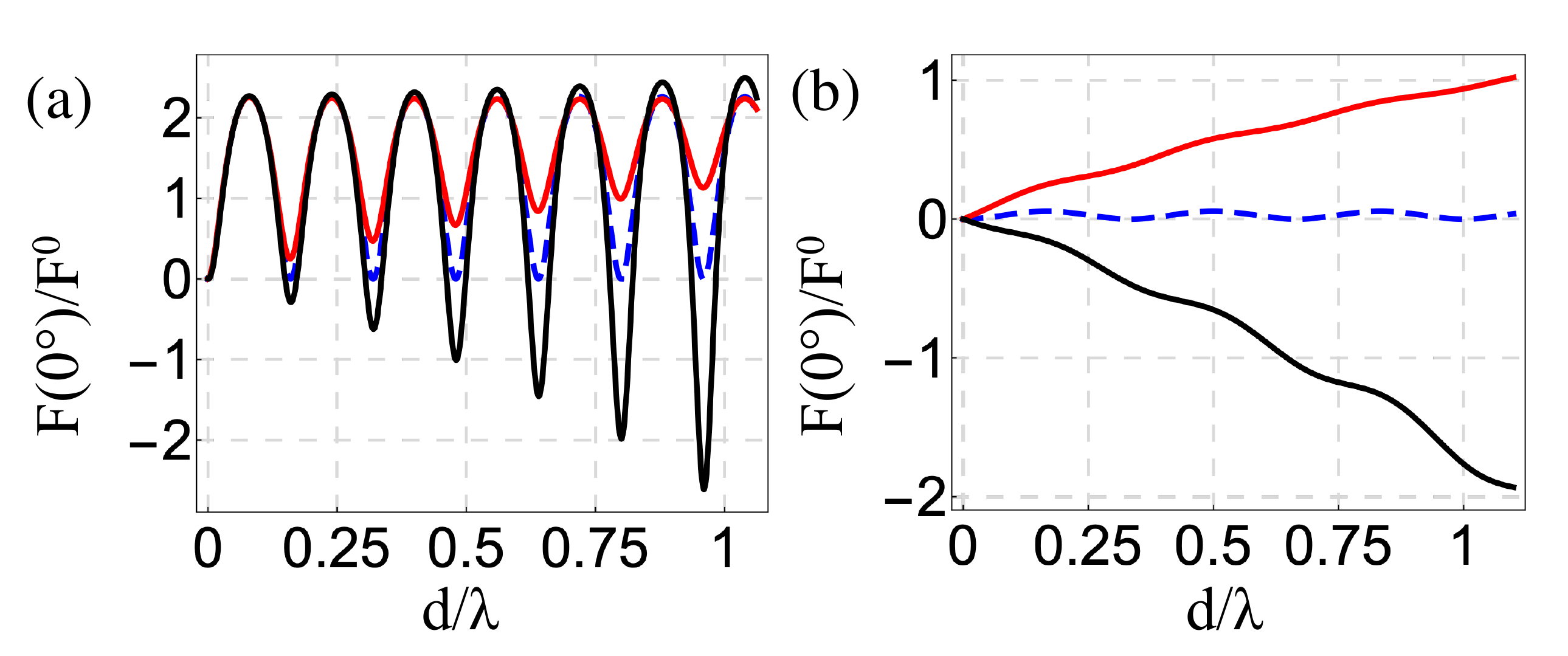}
	\caption{Normalized optical radiation pressure acting on a dielectric slab  by plane wave shining {at a normal incidence} (i. e. we take $\theta=0\degree$) as a function of $d/\lambda$. The refractive index of the slab is (a) $\rm{Re}(n_s) = 3.54$ and (b) $\rm{Re}(n_s) = 1.5$ for different imaginary parts: such as $\rm{Im}(n_s) =- 0.05$  (black), ${\rm Im}(n_s)= 0$ (dashed), and  ${\rm Im}(n_s) = 0.05$ (red).}
	\label{F3} 
\end{figure}

The next step is to find the reflections ($E_{r,\parallel},E_{r,\perp}$)  and transmissions ($E_{t,\parallel},E_{t,\perp}$) coefficients outside the chiral slab. For this purpose, the incident electric
and magnetic fields defined in Eqs. (\ref{Er})-(\ref{Et}), together with expressions for the electromagnetic field inside the chiral slab, obtained by \cite{bassiri}, can further  be written in different regions by applying the appropriate boundary conditions. For instance, one can define:  (i) the reflected and transmitted fields inside the host medium and inside the chiral slab, respectively, at $z=0$, (ii) the reflected and transmitted fields inside the slab and inside the surrounding medium at $z=d$, respectively. This is simple but lengthy process and the all  calculations are rigorously  presented in the Ref. \cite{bassiri}. By following the same procedure we come up with eight equations and eight variables.

For our purpose  we solve  the reflections ($E_{r,\parallel},E_{r,\perp}$)  and transmissions ($E_{t,\parallel},E_{t,\perp}$) coefficients by solving the  system of equations, in the form of $8\times8$ Matrix, (by following the procedure as given in Ref. \cite{bassiri}) by using the Mathematica.  Furthermore,  we consider the both incident plane waves with the same helicities $\sigma_{1}=\sigma_{2}=\sigma=\pm 1$

The radiation pressure on an infinite slab is determined by solving the Maxwell stress tensor through closed Gaussian surfaces (represented by the dashed lines in Fig. \ref{F2}) \cite{Brandon} and given as 
\begin{eqnarray}
 &&\mathbf{F}=\frac{\varepsilon E_{0}^2}{2}\biggl[\biggl(\vert 1-E_{r,\parallel}\vert^2+\vert 1+i\sigma E_{r,\perp}\vert^2-\vert E_{t,\parallel} \vert^2- \vert E_{t,\perp} \vert^2\bigg) \nonumber \\&&cos^2\theta  \,cos^2(kx\sin\theta)
+\biggl( \vert E_{t,\parallel} \vert^2+ \vert E_{t,\perp} \vert^2-\vert 1+E_{r,\parallel}\vert^2-\vert 1\nonumber \\&&-i\sigma E_{r,\perp}\vert^2
\biggr)\biggl\lbrace\sin^2\theta\,\sin^2(kx\sin\theta)-\cos^2(kx\sin\theta)\biggr\rbrace\biggr]\mathbf{\hat{z}}.
\end{eqnarray}

 where $\mathbf{F}=\mathbf{F}(x,\theta)$. In this case, the expressions for the transmission and reflection coefficients are lengthy and complicated, we solve them by using the Mathematica software. The Mathematica file containing  the solution of the problem  can be obtained by contacting the corresponding author. 

\section{Result and discussion} \label{results}

In this section, we present normalized radiation pressure $F/F^0$, where $F^0=\varepsilon E^2_0$, acting on an infinite dielectric (chiral) slab by taking  two linearly (circularly) polarized  plane waves of same polarization and vacuum wavelength $\lambda_0 = 1064 {\rm nm}$. In all examples, we consider refractive index of the slab $n_s =\sqrt{\varepsilon_s\mu_s} $, where $\mu_s = 1$ is the permeability of the slab,  $d$ is the thickness of the slab, and medium around the slab is water with relative refractive index $n_m =1.332$. 

\begin{figure}[hbt!]
	\centering
	\includegraphics[scale=0.650]{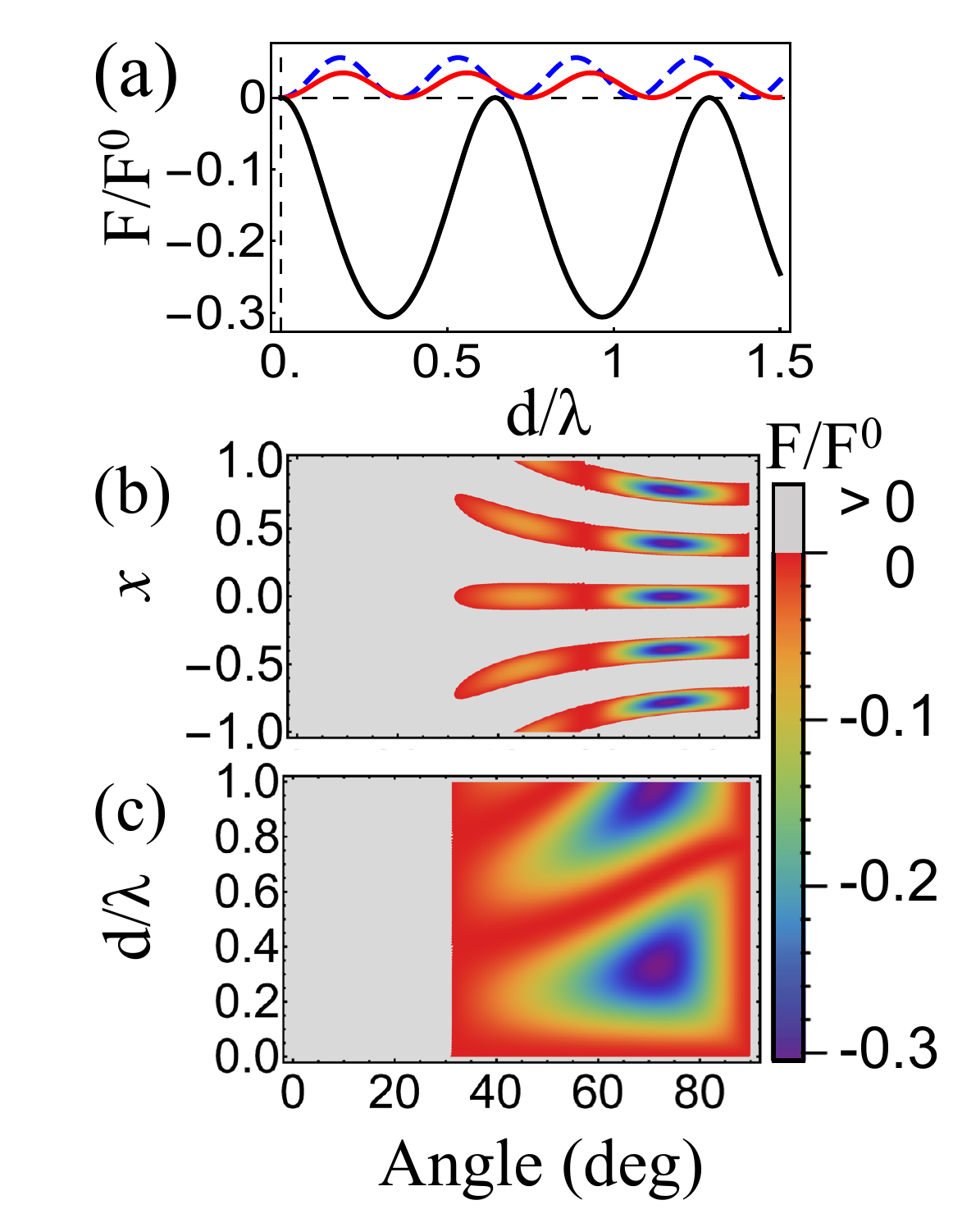}
	\caption{(a) Normalized radiation pressure acting on a dielectric slab by two plane waves impinging at $x= 0$, with  different incident angles: $\theta = 0\degree$ (dashed), $\theta = 20\degree$ (red), $\theta=70\degree $ (black) as a function of $d/\lambda$. Normalized   radiation pressure (b) as function of $\theta$  and traversal position for fixed $d=1.5\lambda$, (c) as function of $\theta$ and $d/\lambda$ for $x=0$, where, the color areas show negative and white area shows positive radiation pressure. }
	\label{F4} 
\end{figure}
First we consider a dielectric slab to elucidate the role of oblique incidence, i.e., angle between the two impinging linearly polarized plane waves, on the radiation pressure by using the analytically derived expression given in Eq. (\ref{force_linear}). For the sake of consistency, one can  get optical force expression on a dielectric slab due to  the normal incidence  by setting $\theta=0$. In this scenario, no interference occurs \ref{F1}(b), and the incident fields exert  a positive  acceleration on the center of mass of the slab by virtue of the law of conservation of momentum. Thus, it  manifests that the radiation pressure by {two incident normal plane waves} is pushing for ${\rm Im}(n_s)\geq 0$ \cite{Mizrahi2010,Wang2015,Li2016}. { {Moreover, for better illustration, we calculate the normalized radiation pressure as a  function of $d/\lambda$ for fixed real refractive index and the results are displayed in Fig. \ref{F3}: (a) ${\rm Re}{(n_s)}=3.54$; and (b) ${\rm Re}{(n_s)}=1.5$. Whereas, in both the cases different values of imaginary part of the refractive index have been used as: $ {\rm Im}(n_s)= 0.05 $ (red line), $ {\rm Im}(n_s)= 0$ (dashed line),  and $ {\rm Im}(n_s)= -0.05 $  (black line).} It is clear that for passive slab, radiation pressure is positive for ${\rm Im}(n_s)\geq 0$}. On the other hand, pressure is negative  when slab is active \cite{Ali2021gain,Ali2022acs,Ali2021FIO} i. e.  ${\rm Im}(n_s)<0$ as demonstrated in Fig. \ref{F3} (black lines). Since there is {no} interference and the pressure is constant over the slab \ref{F1}(b). Thus, the center of mass of the slab  undergoes to a net  pushing force when slab is passive and pulling force when slab is active. Another interesting feature is seen when we increase the thickness of the slab, that is, the force is periodic for a lossless slab because no photons are captured by the lossless slab, however, there is periodically increasing positive pressure for the absorptive case, due to absorptive losses. For active slab, the negative radiation pressure  increases by increasing the $d$ due to the fact that a thicker active slab emits more light than it receives which allows for higher amplification of the forward momentum, consequently increasing the negative radiation pressure. 

It is worth mentioning that the slab with larger ${\rm Re}(n_s)$ presents larger  reflection, therefore, it needs a thick active slab (i.e., large emission requires to overcome the radiation  losses) to  revert the direction of radiation pressure. For instance, in the Fig \ref{F3}(a)  ($n_s = 3.54-0.05i$) the radiation pressure changes sign at $d/\lambda = 0.2$ and in Fig.  \ref{F3}(b)  ($n_s = 1.5-0.05i$)  and the pressure becomes negative for small thickness.

The situation changes dramatically, when the two plane waves make an oblique incidence at an angle $\theta>0\degree $,  which lead to an interference effect between the sources as shown in \ref{F1}(c). As a result, the transmitted  and  reflected lights give local  maxima and minima along the x-direction and hence the corresponding radiation pressure over the surface will provide periodic locally positive and negative interference pattern \ref{F1}(b). In order to observe local positive  and negative pressure on the slab we separate the both sources by an angle {$2\theta$} as shown  in Fig. \ref{F1}.

In Fig. \ref{F4}, we consider two linearly polarized plane waves impinging on a dielectric infinite slab of refractive index $n_s= 1.5$.  Fig. \ref{F4}(a) presents the radiation pressure as a  function of $d/\lambda$ for different angles: $\theta = 0\degree$ (dashed), $\theta = 20\degree$ (red), and $\theta = 70\degree $ (black) for a fixed transversal position $x=0$. It is clear that for $\theta= 0$ and $\theta=20 \degree$ radiation pressure is always positive which means that scattered momentum is larger over the slab. However, for the large angle the radiation pressure becomes negative which appears due to the increase of  forward momentum than the backward momentum and hence pressure is locally negative. This locally  positive and negative radiation pressure  pattern appears along the $x$-axis of infinite slab, which is shown in Fig. \ref{F4}(b).  It is clearly shown that for $\theta> 30\degree$ the dielectric slab undergoes to a local pulling and pushing force densities. The crossover between the positive and negative radiation pressure is  shown in Fig. \ref{F4}(c), where optical radiation pressure is calculated  as a function of  angle and $d/\lambda$  for fixed $x=0$. It indicates that the slab at the point of incidence experiences   a positive radiation pressure    $(F/F^0>0)$
for angle $0 \degree\leq\theta<30\degree$ and local negative radiation pressure $(F/F^0<0)$  for  $30 \degree<\theta<90\degree$. 

\begin{figure}[hbt!]
	\centering
	\includegraphics[scale=0.40]{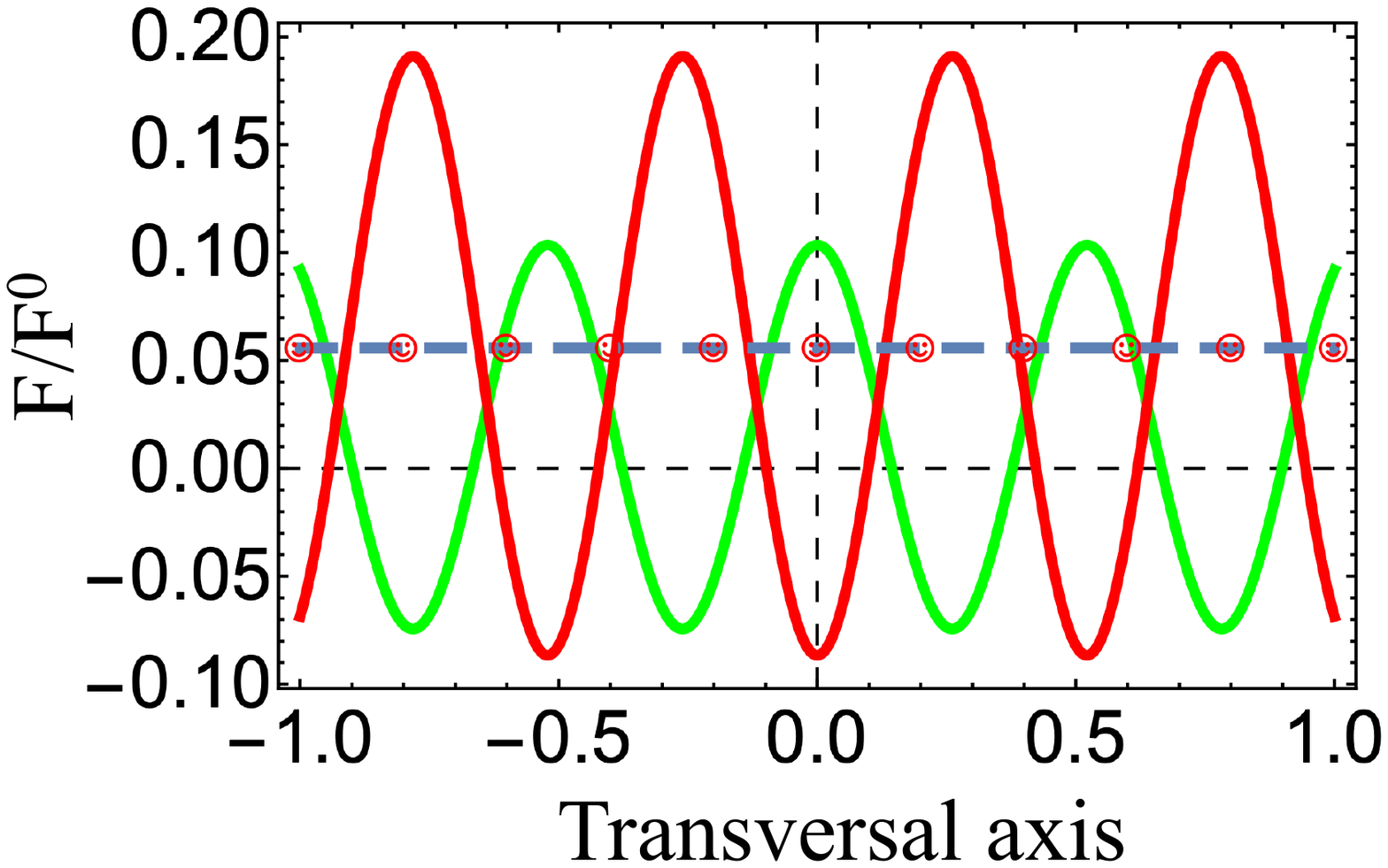}
	\caption{Normalized radiation pressure by two left circularly polarized plane waves incident at $\theta=50\degree $ (green and red lines) acting on chiral slab as a function of  transversal position for different chirality parameter: such as $\kappa = +0.3$ (red), $\kappa = -0.3$ (green).   The dashed (emoji) line is calculated for normal incident (i. e. $\theta = 0\degree$) for chirality parameter  $\kappa=-0.3$ ($\kappa=0.3$). The thickness of the slab is fixed $d = 1.5 \lambda$}.
	\label{F5} 
\end{figure}

For many applications, natural chiral materials such as proteins, organic compounds,  nucleic acids \cite{Corradini2007}, {and} chiral metamaterials and  metasurfaces are widely used in biological applications.  In this context,  if the slab is made of such a chiral material, then a TE plane wave exerts an identical radiation pressure on both chiral enantiomers with opposite chirality parameter $\kappa$, as expected from symmetry \cite{Ali2021}.  It is due to fact that the TE plane wave reads the same refractive index regardless of $\kappa$ and hence cannot create the specific positive and negative radiation pressure intervals depending slightly on $\kappa$. 
To circumvent this issue we come up with a new strategy, i.e.,  by shining two circularly polarized plane waves on infinite  chiral slab as shown in Fig \ref{F2}. Now each polarization reads distinct refractive index $n_s= \sqrt{\epsilon_s}+\sigma \kappa$, thus exerted pressure on the chiral slab is  obliged to  be distinct.

\begin{figure}[hbt!]
	\centering
	\includegraphics[scale=0.450]{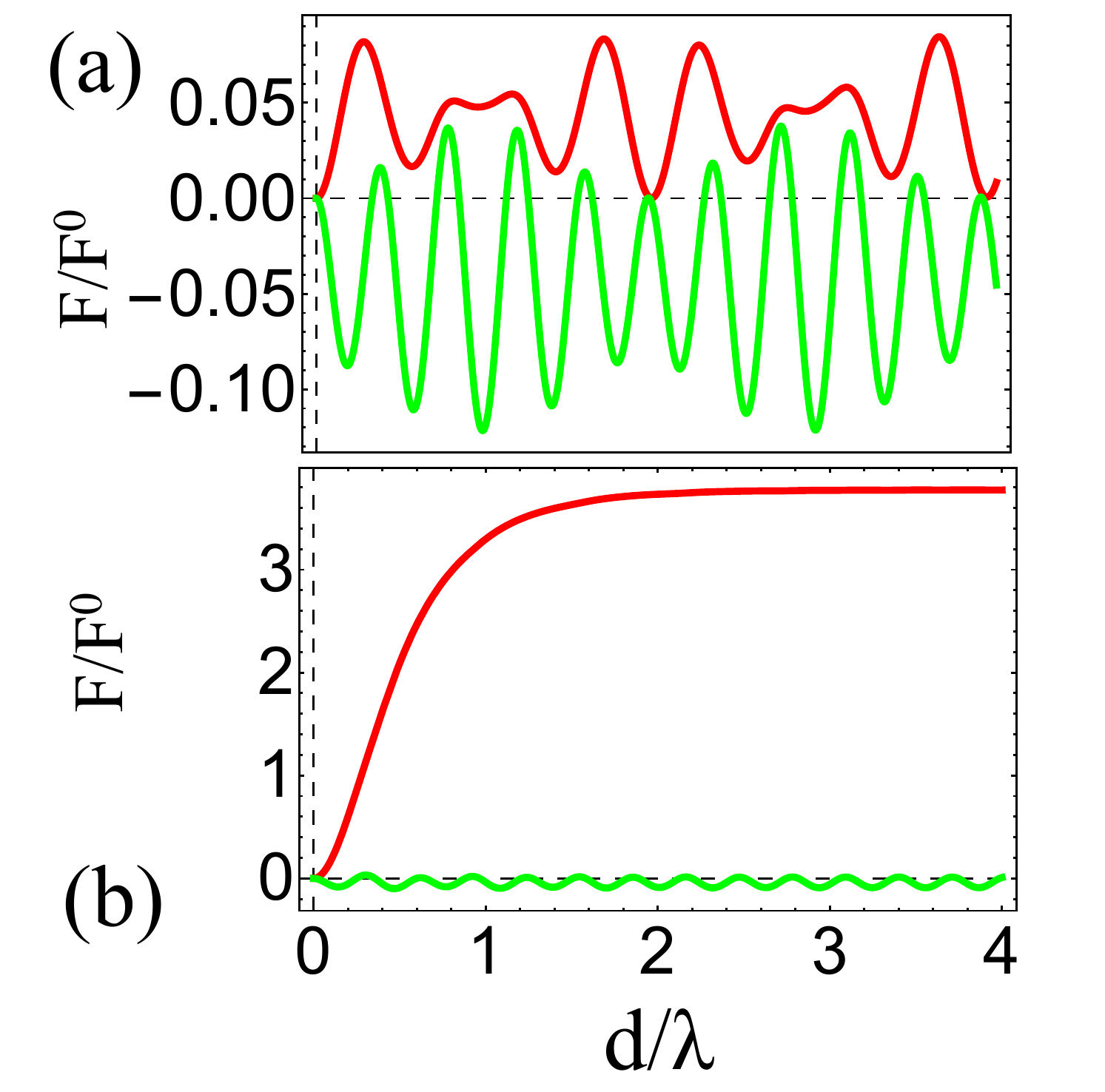}
	\caption{Radiation pressure by two LCP plane waves impinging at an angle $\theta=50\degree $ acting on dielectric slab as a function of $d/\lambda$ (a) for chirality parameter $\kappa = -0.2$ (red) and $\kappa = 0.2$ (green) (b)   $\kappa = -0.5$ (red) and $\kappa = 0.5$ (green).}
	\label{F6} 
\end{figure}
In Fig. \ref{F5} we demonstrate the normalized radiation pressure acting  on a chiral slab of thickness $d$ refractive index $n_s = 1.5$  as a function of the transversal axis for chirality parameter: $\kappa= - 0.3$ and $\kappa= 0.3$ at different angles. First, we select normal incidence $\theta = 0\degree$ and the radiation pressure is calculated for  $\kappa = 0.3$ (dashed line) and  $\kappa = -0.3$ (emoji). It  suggests that the radiation pressure is constant along the tranversal direction, as expected (due to the vanishing of interference) but it also manifests that the radiation pressure does not depend on chirality for normal incidence, see dashed and emoji lines in  Fig. \ref{F5}.  Thus, the electromagnetic field impinges at angle i. e. $\theta = 50\degree$, thereby, the interference comes into play. The chiral slab undergoes a local positive and negative pressure, which explicitly depends on the chirality parameters.
 For instance, left circularly polarized light exerts negative (positive) radiation pressure on the chiral slab with $\kappa=0.3(\kappa =-0.3)$ at origin (point of incidence), which propagates along the transverse direction.  
The direction and magnitude of the {radiation pressure} can be changed by changing the angle {of the impinging plane waves} and the thickness of the slab.

In Fig. \ref{F6}, we unveil the role of chirality by calculating the radiation pressure as a function of $d/\lambda$  for different chirality parameters: Fig. \ref{F6}(a) $\kappa= 0.2$ (green)  $\kappa= -0.2$ (red) Fig. \ref{F6}(b) $\kappa= 0.5$ (green)  $\kappa= -0.5$ (red) for fixed left circularly polarized  light sources impinging at  $x=0$. {For this particular configuration, chiral slab with  $\kappa>0 $ (as the same handedness as of the light sources)  undergoes to a negative local radiation pressure  when the handedness of chirality is reverted (i.e. $\kappa<0 $), then the pressure is positive.} It can be associated with the well know fact that a chiral medium favours the transmission of light if the handed of the chiral is same as the handed of the polarization and on contrast optimizes the reflection for opposite handedness.

This effect is elaborated in Fig \ref{F6}(b) by taking large chirality parameters, where radiation pressure corresponding to $\kappa = - 0.5$ presents a maximum positive pressure because the maximum of the incident signal gets reflected due to opposite handedness and the radiation pressure corresponding to $\kappa = 0.5$ (green line)  provides a maximum negative pressure, as expected by the symmetry consideration. This {pressure} can be reversed by changing the polarization of the incident beam.  Indeed, the radiation pressure is invariant by changing both $\kappa$ and $\sigma$ {in the same way}.

\section{Conclusions} \label{discussion}
 In conclusion, we have developed, for the first time, a theoretical framework to calculate the local radiation pressure on an infinite dielectric (chiral) slab illuminated by a superposition of linearly (circularly) polarized plane waves. In this prospective, we have derived explicit analytical results for the optical radiation pressure acting on an infinite slab by following two approaches. In the first approach,  we have solved the standard Maxwell stress tensor to calculate the total optical radiation pressure by incorporating the total fields on each side of the slab. Consequently, the slab experiences a total local positive or negative radiation pressure.  In the second approach, we have calculated the  Maxwell stress tensor at each interface of the slab separately and then obtained the explicit expressions for the radiation pressure on respective interfaces. However, subsequently, the superposition of the radiation pressures of the both interfaces  provides the total radiation pressure on the slab. Moreover, we have not only verified that the final results obtained by the both  approaches are in good agreement but we also have shown, by considering the limiting cases (for instance, by taking the incident angle equal to zero), that our analytical results perfectly agree with the well-known results of the radiation pressure acting on a slab due to normal impinging field. In addition to the analytical work, we have considered realistic  material parameters to numerically validate our findings. These results have shown that the local radiation pressure strongly depends on the incident angle, polarization of the light source and the nature of the material. For instance, for a normal incidence of the field, the radiation pressure is constant all over the slab, whereas, for an oblique incidence it varies locally along the transversal direction, which occurs due to the interference effects. In the case of chiral slab, chirality parameter also plays a key role to govern the local minima and maxima of the radiation pressure, which are indeed consistent with the momentum and energy conservation laws. 
 At last but not the least,  we hope that this approach paves the path for calculating the radiation pressure that can be used, in conjunction with numerical simulations, to yield the distribution of fields, as well as the resulting interference patterns and corresponding local extension and relaxation of soft matter, cell membranes in diverse biological systems of practical interest.

\appendix
\section*{Appendix}

\section{Force at the interfaces}

In this appendix we present the principal steps to obtain the radiation pressure due to two non-collinear plane waves on each interfaces of the slab as shown in Fig. \ref{F1}. 

In order to calculate the radiation pressure $\mathbf{F}(z=0,x,\theta)$ at the interface $z=0$,  we take the flux through the Gaussian surface at $z=0$ and  the resultant $\mathbf{F}(z=0,x,\theta)$ is expression in terms of  the Maxwell stress tensor as

\begin{equation}
\mathbf{F}(z=0,x,\theta) = \frac{1}{2}\Re\lbrace \mathbf{\hat{z}}\cdot \overleftrightarrow{T}(z=0^{-})- \mathbf{\hat{z}}\cdot\overleftrightarrow{T}(z=0^{+})\rbrace. \label{pressure}
\end{equation}

In this case the electromagnetic fields inside the slab  can be expressed by two types of plane waves propagating with positive and negative $z$ components of wave vectors 

\begin{equation}
\mathbf{E}_{w,k_z>0}=2E_{w}^{(1)}cos(k_{s}x\sin\theta_s)e^{i k_sz\cos\theta_s}e^{-i \omega t}\mathbf{\hat{y}}  \label{fieldin1}
\end{equation}
and 
\begin{equation}
\mathbf{E}_{w,k_z<0}=2E_{w}^{(2)}cos(k_{s}x\sin\theta_s)e^{-i k_sz\cos\theta_s}e^{-i \omega t}\mathbf{\hat{y}}, \label{fieldin2}
\end{equation}
respectively, where $k_{s}$ is the wave vector inside slab. In addition, 

\begin{equation}
E_{w}^{(1)}=\frac{E_{0}t_{m,s}}{1-r_{s,m}^2\,e^{i2\beta}}
\end{equation}
and 
\begin{equation}
E_{w}^{(2)}=\frac{E_{0}t_{m,s}r_{s,m}e^{i\beta}}{1-r_{s,m}^2\,e^{i2\beta}}.
\end{equation}

The total electric field is given by $\mathbf{E}_{total,w}=\mathbf{E}_{w,k_z>0}+\mathbf{E}_{w,k_z<0}$. Furthermore, the total  electric fields at the both sides of the Gaussian interface $z=0$ are evaluated, $\mathbf{E}_{total}(z=0^{-})=\mathbf{E}_{in,total}(z=0 ^{-})+\mathbf{E}_{r,total}(z=0^{-})$ and $\mathbf{E}_{total}(z=0^{+})=\mathbf{E}_{total,w}(z=0^{+})$. By setting these fields along with   Eqs. \ref{Einp}-\ref{Erp} and \ref{fieldin1}-\ref{fieldin2}  in Eq.\ref{tensor} we obtain the radiation pressure acting on at the interface $z=0$:

%\[
%\mathbf{F}(z=0,x,\theta) = \varepsilon E_{0}^2\biggl[\biggl(\vert 1+r_{TE}\vert^2+\vert 1 - r_{TE}\vert^2 \cos^2\theta\biggr)\,\cos^2(kx\sin\theta)- \]\[ \vert 1+r_{TE} \vert^2\,\sin^2\theta\,\sin^2(kx\sin\theta)\biggr]\mathbf{\hat{z}}
%-\,\varepsilon_{s} E_{0}^2\biggl[\biggl(\vert t_{w}+r_{w}e^{i\beta}\vert^2+\vert t_{w} - r_{w}e^{i\beta}\vert^2 \cos^2\theta_{s}\biggr)\times \]\begin{equation}\,\cos^2(k_{s}x\sin\theta_{s})- \vert t_{w}+r_{w}e^{i\beta} \vert^2\,\sin^2\theta_{s}\,\sin^2(k_{s}x\sin\theta_{s})\biggr]\mathbf{\hat{z}}, \label{forcezero}
%\end{equation}

\begin{eqnarray}
&&\mathbf{F}(z=0,x,\theta) = \varepsilon E_{0}^2\biggl[\biggl(\vert 1+r_{TE}\vert^2+\vert 1 - r_{TE}\vert^2 \cos^2\theta\biggr) \nonumber\\&&\cos^2(kx\sin\theta)- \vert 1+r_{TE} \vert^2\,\sin^2\theta\,\sin^2(kx\sin\theta)\biggr]\mathbf{\hat{z}}\nonumber\\&&
-\,\varepsilon_{s} E_{0}^2\biggl[\biggl(\vert t_{w}+r_{w}e^{i\beta}\vert^2+\vert t_{w} - r_{w}e^{i\beta}\vert^2 \cos^2\theta_{s}\biggr) \nonumber\\&&\cos^2(k_{s}x\sin\theta_{s})- \vert t_{w}+r_{w}e^{i\beta} \vert^2\,\sin^2\theta_{s}\,\sin^2(k_{s}x\sin\theta_{s})\biggr]\mathbf{\hat{z}}, \nonumber \\\label{forcezero}
\end{eqnarray}

where
\begin{equation}
t_{w}=\frac{t_{m,s}}{1-r_{s,m}^2e^{i2\beta}}
\end{equation}

and

\begin{equation}
r_{w}=\frac{t_{m,s}r_{s,m}}{1-r_{s,m}^2e^{i2\beta}}.
\end{equation}

Similarly, the radiation pressure $\mathbf{F}(z=d,x,\theta)$ at the interface $z=d$ is given by

\begin{equation}
\mathbf{F}(z=d,x,\theta) = \frac{1}{2}\Re\lbrace \mathbf{\hat{z}}\cdot \overleftrightarrow{T}(z=d^{-})- \mathbf{\hat{z}}\cdot\overleftrightarrow{T}(z=d^{+})\rbrace. 
\end{equation}

The total electric fields in the both sides of the gaussian surface are given by $\mathbf{E}_{total}(z=d^{-})=\mathbf{E}_{total,w}(z=d^{-})$ and $\mathbf{E}_{total}(z=d^{+})=\mathbf{E}_{t,total}(z=d^{+})$, and the radiation pressure is obtained setting the Eqs. \ref{Etp} and \ref{fieldin1}-\ref{fieldin2} in Eq.\ref{tensor}

\begin{eqnarray}
&&\mathbf{F}(z=d,x,\theta)= -\varepsilon E_{0}^2\vert r_{TE} \vert^2 \biggl[\biggl(1+ \cos^2\theta\biggr)\,\cos^2(kx\sin\theta)\nonumber\\&&- \sin^2\theta\,\sin^2(kx\sin\theta)\biggr]\mathbf{\hat{z}}
+\,\varepsilon_{s} E_{0}^2\biggl[\biggl(\vert t_{w}e^{i\beta}+r_{w}\vert^2\nonumber\\&&+\vert t_{w}e^{i\beta} - r_{w}\vert^2 \cos^2\theta_{s}\biggr)\,\cos^2(k_{s}x\sin\theta_{s})-\vert t_{w}e^{i\beta}+r_{w} \vert^2\nonumber\\&& \sin^2\theta_{s}\sin^2(k_{s}x\sin\theta_{s})\biggr]\mathbf{\hat{z}}.\label{forced}
\end{eqnarray}

Finally, the total radiation pressure acting the slab is written as
\begin{equation}
   \mathbf{F}= \mathbf{F}(x,\theta)=\mathbf{F}(z=0,x,\theta)+\mathbf{F}(z=d,x,\theta) \label{totalforcer}
\end{equation}
It is clear from Eqs. (\ref{forcezero}) and  (\ref{forced}) that the net momentum  flux inside the slab is conserved, and  the total radiation pressure on the surface is given in Eq.  (\ref{totalforcer}) which is the same as we have presented in Eq. (\ref{force_linear}). Thus, by analysing the Eq. (\ref{totalforcer}), it clear that one can not only recover Eq. (\ref{force_linear}) for the case of losses slab but also recover Eq. (\ref{F_norm}), for normal incidence  even considering the case when  imaginary part of the refractive index is different than zero $Im(n_s)\neq 0$.

\section*{Funding}  We thank G. Wiederhecker, F. A. Pinheiro, F. S. S. Rosa and P. A. M. Neto for inspiring discussion. 
This work is partially supported by 
 Fundac\~ao de Amparo a   Pesquisa do Estado de S\~ao Paulo  (FAPESP) (2020/03131-2). 
 The authors declare no conflicts of interest.


\begin{thebibliography}{99} 

\bibitem{Ashkin1970} A. Ashkin, \textit{Acceleration and trapping of particles by radiation pressure}, Phys. Rev. Lett. {\bf 24}, 156 (1970).

\bibitem{ashkin1986}  A. Ashkin, J. M. Dziedzic, J. E. Bjorkholm, and S. Chu,  \textit{Observation of a single-beam gradient force optical trap for dielectric particles,}   Opt. Lett. {\bf 11}, 288 (1986).


\bibitem{Gieseler2020} J. Gieseler,  J. R. Gomez-Solano,  et al., \textit{Optical Tweezers: A Comprehensive Tutorial from Calibration to Applications,}
Adv. Opt. Photon. \textbf{13}, 74-241 (2021).

\bibitem{Polimeno2018} P. Polimeno, A. Magazz\'u, M. A. Iat\'i, et al.,Optical tweezers and their applications, J. Quant. Spectrosc. Radiat. Transf. \textbf{218}, 131-150 (2018).



\bibitem{ashkin2006}  A. Ashkin,  \textit{Optical trapping and manipulation of neutral particles using lasers}: A reprint volume with commentaries (World Scientific, Singapore, 2006).


\bibitem{Friese1998} M. E. J. Friese, T. A. Nieminen, N. R. Heckenberg, and H. Rubinsztein-Dunlop, \textit{Optical alignment and spinning of
laser-trapped microscopic particles,}  Nature {\bf 394}, 348 (1998).

\bibitem{grier2003} D. G. Grier,  \textit{A revolution in optical manipulation}, Nature {\bf 424},  810  (2003).

\bibitem{Lin2014}  J. Lin and Y. Q. Li, \textit{Optical trapping and rotation of airborne absorbing particles with a single focused laser beam,} Appl. Phys. Lett. {\bf 104}, 101909 (2014).

\bibitem{Ali2020Tailor} R. Ali, F. A. Pinheiro, R. S. Dutra, and P. A. Maia Neto, \textit{Tailoring optical pulling forces with composite microspheres}, Phys. Rev. A {\bf 102}, 023514 (2020).

\bibitem{Ali2021} R. Ali,  R. S. Dutra, F. A. Pinheiro, and  P. A. Maia Neto,  \textit{Enantioselection and chiral sorting of single microspheres using optical pulling forces }, Opt. Lett. {\bf 46}, 1640 (2021).


\bibitem{chan2011}  J. Chen, J. Ng,  Z. F. Lin, and C. T. Chan, \textit{Optical pulling  force,}  Nat. Photonics { \bf{5}}, 531 (2011).

\bibitem{li2019}  X. Li, J.  Chen, Z. Lin, J. Ng.  \textit{Optical pulling at macroscopic distances}, Sci. Adv.  \textbf{5}, 7814 (2019).
\bibitem{Shvedov2014} V.  Shvedov, A. R.  Davoyan,  N. Engheta and  W. Krolikowski, \textit{A long-range polarization-controlled optical tractor beam}, Nat. Photonics \textbf{ 8},  846  (2014).

\bibitem{dogariu2011} S. Sukhov and A. Dogariu, \textit{Negative nonconservative forces: optical tractor beams for arbitrary objects,} Phys. Rev. Lett. \textbf{107}, 203602 (2011).

\bibitem{Ali2018} R. Ali, F. A. Pinheiro, F. S. S. Rosa, R. S. Dutra, P. A. Maia Neto,  \textit{Optimizing optical tweezing with directional scattering in composite microspheres}, Phys. Rev. A {\bf 98 },  053804 (2018).



\bibitem{Stephen2007} S. H. Simpson, and and S. Hanna, {\it Optical trapping of spheroidal particles in Gaussian beams} , J. Opt. Soc. Am. A {\bf 24}, 430-443 (2007).

\bibitem{Wiley2007}  B. J. Wiley, Y.  Chen,  J. M. McLellan,  Y. Xiong,  Z.-Y. Li,   D. Ginger, and  Y. Xia, {\it Synthesis and Optical Properties of Silver Nanobars and Nanorice, }  Nano Letters, 7, 1032-1036 (2007).

\bibitem{Gauthier1999} R. C. Gauthier, M.  Ashman, and C. P. Grover, \textit{Experimental confirmation of the optical-trapping properties of cylindrical objects,} Appl. Opt. \textbf{38}, 4861-4869 (1999).



\bibitem{Ali2020} R. Ali, F. A. Pinheiro, F. S. S. Rosa, R. S. Dutra, P. A. Maia Neto, \textit{ Enantioselective manipulation of chiral nanoparticles using optical tweezers},  Nanoscale {\bf 12}, 5031 (2020).
\bibitem{Ali2020josa}  R.  Ali, F. A. Pinheiro, R. S. Dutra, F. S. S. Rosa, and P. A. Maia Neto, \textit{Probing the optical chiral response of single nanoparticles with optical tweezers}, J. Opt. Soc. Am. B \textbf{37}, 2796-2803 (2020).

\bibitem{Canaguier2013} A. Canaguier-Durand, J. A. Hutchison, C. Genet, and T. W. Ebbesen, \textit{Mechanical separation of chiral dipoles by chiral light}, New J. Phys. \textbf{15}, 123037 (2013).

\bibitem{Kun2014}  K.  Ding, J. Ng, L. Zhou, and C. T. Chan, \textit{ Realization of optical pulling forces using chirality,} Phys. Rev. A {\bf 89}, 063825  (2014).

\bibitem{Mansuripur2005} M. Mansuripur, \textit{Radiation pressure and the linear momentum of the electromagnetic field}, Opt. Express \textbf{12}, 5375-5401 (2004).

\bibitem{Loudon2010} R. Loudon and S. M. Barnett, \textit{Theory of the radiation pressure on dielectric slabs, prisms and single surfaces}, Opt. Express \textbf{14}, 11855-11869 (2006).
{ \bibitem{Alaee2018} R. Alaee, J. Christensen, and M. Kadic, \textit{Optical Pulling and Pushing Forces in Bilayer PT -Symmetric Structures}, Phys. Rev. Appl. \textbf{9}, 014007 (2018)}


\bibitem{Corradini2007}, R. Corradini, S. Sforza, T. Tedeschi, and R. Marchelli, \textit{Chirality as a tool in nucleic acid recognition: Principles and relevance in biotechnology and in medicinal chemistry}, Chirality \textbf{19}, 269, (2007)

\bibitem{Basiri} A. Basiri, X. Chen, J.  Bai,  et al. {\it Nature-inspired chiral metasurfaces for circular polarization detection and full-Stokes polarimetric measurements,} Light Sci. Appl. \textbf{8}, 78 (2019).

\bibitem{Lepeshov2020} S.  Lepeshov, and A. Krasnok, \textit{Virtual optical pulling force,} Optica, \textbf{7}, 1024-1030 (2020).

\bibitem{Mizrahi2010} A. Mizrahi, and Y. Fainman, \textit{Negative radiation pressure on gain medium structures}, Opt. Lett. \textbf{35}, 3405-3407 (2010).
\bibitem{Wang2015} M. Wang, H. Li, D. Gao, L. Gao, J. Xu, and Cheng-Wei Qiu, \textit{Radiation pressure of active dispersive chiral slabs,} Opt. Express \textbf{23}, 16546-16553 (2015).

\bibitem{Li2016} G. Li, M. Wang, H. Li, M. Yu, Y. Dong, and Jun Xu, \textit{Wave propagation and Lorentz force density in gain chiral structures}, Opt. Mater. Express \textbf{6}, 388-395 (2016).



 \bibitem{Ali2021gain} R. Ali, R. S. Dutra, F. A. Pinheiro and P. A. Maia Neto,  Gain-assisted optical tweezing of plasmonic and large refractive index microspheres,  J. Opt. \textbf{23}, 115004, (2021).
 



\bibitem{Born}  M. Born, and  E. Wolf. \textit{Principles of optics: electromagnetic theory of propagation, interference and diffraction of light}, Elsevier, (2013).

\bibitem{Brandon} A. B. Kemp, M. T.  Grzegorczyk, and  J. A.  Kong,  \textit{Ab initio study of the radiation pressure on dielectric and magnetic media}, Opt. Express  \textbf{13},  9280-9291 (2005).


\bibitem{Pfeifer2007} R. N. C. Pfeifer, T. A. Nieminen, N. R. Heckenberg, and H. Rubinsztein-Dunlop,  Momentum of an electromagnetic wave in dielectric media, Rev. Mod. Phys. \textbf{79}, 1197 (2007).
\bibitem{wang2014} S. B. Wang and C.T. Chan, \textit{Lateral optical force on chiral particles near a surface}, Nat. Commun. \textbf{5}, 3307 (2014).
\bibitem{Bohren} C. F. Bohren and D. R. Huffman, \textit{Absorption and Scattering of Light by Small Particles}, (Wiley, New York, 1983).
\bibitem{Ali2020Theory} R. Ali, R. S. Dutra, P. A.  Pinheiro, F. S. S. Rosa, and P. M. A. Neto, \textit{Theory of optical tweezing of dielectric microspheres in chiral host media and its applications}, Sci. Rep. \textbf{10}, 16481 (2020).
    	


\bibitem{Ali2021FIO} R. Ali, T. P. M. Alegre, F. A. Pinheiro, and G. S. Wiederhecker, "Enantioselective optical forces of gain functionalized core-shell chiral nanoparticles," in Frontiers in Optics + Laser Science 2021, C. Mazzali, T. (T.-C.) Poon, R. Averitt, and R. Kaindl, eds., Technical Digest Series (Optica Publishing Group, 2021), paper FTh4E.2.
 
 \bibitem{Ali2022acs} R. Ali,  Tunable anomalous scattering and negative asymmetry parameter in a gain-functionalized low refractive index sphere, ACS Omega  {\bf{7}}, 2170–2176 (2022). 



\bibitem{bassiri} S. Bassiri, C. H. Papas, and N. Engheta, \textit{ Electromagnetic wave propagation through a dielectric chiral interface and through a chiral slab}, J. Opt. Soc. Am. A, \textbf{ 5},   1450-1459, (1988).



 








\end{thebibliography}
\end{document}